\documentstyle[12pt,fleqn]{article}
\newcommand{\Tr}{\mathop{\rm Tr}\nolimits}
\newcommand{\e}{\mathop{\rm e}\nolimits}
\newcommand{\ts}{\textstyle}
\newcommand{\ds}{\displaystyle}
\newcommand{\ints}{\int\limits}
\newcommand{\nn}{\nonumber\\}
\newcommand{\eps}{\varepsilon}
\begin{document}
\begin{titlepage}

\begin{center}
\large\bf
Chiral phase transitions in strong chromomagnetic fields at finite
temperature and dimensional reduction
\end{center}
\vspace{0.5cm}
\renewcommand{\thefootnote}{\fnsymbol{footnote}}
\begin{center}
D. Ebert\footnotemark[1],
V.~Ch.~Zhukovsky\footnotemark[1]\footnotemark[2]\\
{\sl Institut f\"ur Physik, Humboldt--Universit\"at zu Berlin,\\
Invalidenstra{\ss}e 110, D--10115 Berlin, Germany}
\end{center}

\begin{abstract}
Dynamical fermion mass generation in external chromomagnetic fields
is considered at non--zero temperature. The general features
of dynamical chiral symmetry breaking ($D\chi SB$) are investigated for several
field configurations in relation to their symmetry properties and the form
of the quark spectrum. According to the fields, there arises dimensional
reduction by one or two units. In all cases there exists $D\chi SB$ even at
weak quark attraction, confirming the idea about the dimensional insensitivity
of this mechanism in a chromomagnetic field.
\end{abstract}

\vspace{0.3cm}
\setcounter{footnote}{1}
\noindent\footnotetext{\noindent Supported by
{\it Deutsche Forschungsgemeinschaft} under contract 436 RUS 113-29.}
\setcounter{footnote}{2}
\noindent \footnotetext{\noindent On leave of absence from the
Faculty of Physics, Department of Theoretical
Physics, Moscow State University, 119899, Moscow, Russia.}
\vfill
\end{titlepage}

%%%%%%%%%%%%%%%%%%%%%%%%%%%%%
\renewcommand{\thefootnote}{\arabic{footnote}}
\setcounter{footnote}{0}
\setcounter{page}{1}
%%%%%%%%%%%%%%%%%%%%%%%%%%%%%

\section*{1. Introduction}

As is well known, the high energy region (short distances) of strong
interactions can be considered perturbatively in the framework of QCD due to
its property of asymptotic freedom. On the other hand, in order to study nonperturbative
phenomena at low energies (large distances) like
the QCD vacuum with gluon and quark condensates [1] and the hadronization process
analytically, various approximate methods and effective models have been proposed.
One approximate possibility to simulate a realistic
gluon condensate is to introduce an appropriate external (abelian or
non--abelian) chromomagnetic field [2]. Exact solutions of particle equations in external
field models allow in particular for explicit analytical calculations and estimates of
various nonperturbative effects, demonstrating nonanalytic dependence of them on
field intensities [3--6]. There exist also simple field--theoretic models like the
Nambu--Jona--Lasinio (NJL) chiral quark model which has been successfully used to
describe $D\chi SB$, bosonization and low energy behavior of mesons (see e.g. [7] and
references therein). In particular, for a QCD--motivated NJL--model with gluon condensate
and finite temperature, it was shown that a weak gluon condensate plays
a stabilizing role for the behavior
of the constituent quark mass, the quark condensate, meson masses and coupling
constants for varying temperature [8].

The effect of dynamical chiral symmetry breaking ($D\chi SB$) under the influence of
a magnetic field, resulting in a
fermion mass generation, has recently attracted much attention [9, 10]. In the framework of four--
fermion models it
has been shown that a constant magnetic field serves as a catalyzing factor
in the fermion mass generation even under conditions, when the interaction
between fermions is rather weak [11--12]. Moreover, in a recent paper [13] it has been
argued that $D\chi SB$ does not take place in an external
axial--symmetric chromomagnetic field in $D=3+1$ dimensions when attraction
between fermions is weak. Thus, this result seems to contradict the idea about
the dimensional insensitivity of this mechanism in a chromomagnetic field.

The purpose of the present letter is to investigate the phenomenon of $D\chi SB$
for various external chromomagnetic fields like non--abelian axial--symmetric and
rotational--symmetric as well as abelian fields. In particular, we will show that in all cases
even for weak coupling of quarks the
$D\chi SB$ does exist confirming the idea about its dimensional insensitivity
and that it is related to an effective dimensional reduction.  Furthermore, we
will find a simple relation between symmetry properties of external fields, the
degeneracy of quark energy spectra and the phenomenon of dimensional
reduction. The latter effect leads to a nonanalytic
logarithmic dependence of the quark condensate on the field strength in the strong field
limit. Finally, we shall consider the effect of finite temperature and
show that in the strong field limit there exists a finite critical
temperature at which a phase transition takes place and chiral symmetry
is restored in both abelian and non--abelian models of the gluon condensate.
In particular, there arises an interesting relationship  $T_c=Cm(0)$
between the critical temperature and the zero temperature fermion mass $m(0)$, with a
universal constant $C$ for different fields.

\section*{2. Quark condensates in external fields}

\subsection*{2.1 General definitions}

Let us consider an $SU(N_f)$ flavor--symmetric model of quarks $q_i$ moving in an
external chromomagnetic field of the color group $SU(3)_c$. The Lagrangian takes
the following form (for convenience, we choose Euclidean
D--dimensional space--time with $it=x_D$)

\begin{eqnarray}
{\cal L}&=&\sum_{i=1}^{N_f}\bar{q_i}(\gamma _\mu \nabla _\mu +m_i)q_i,
\label{1}
\end{eqnarray}
where $m_i=m$ are (equal) masses of quarks, $\nabla _\mu =\partial _\mu
-ig\frac 12\lambda _aA_\mu ^a$ is the covariant derivative of quark fields
in the constant external fields $F_{\mu \nu }^a=\partial _\mu A_\nu
^a-\partial _\nu A_\mu ^a+gf_{abc}A_\mu ^bA_\nu ^c$, determined by the
potentials $A_\mu ^a\left( a=1,...,8\right) $. Next, introduce the vacuum functional $Z_q$ of
quark fields,

\begin{eqnarray}
Z_q &=&\int dqd\bar{q}\exp [-\int d^Dx{\cal L}]=\prod_{i=1}^{N_f}\det (\gamma_\mu \nabla_\mu
+m)=\nn
&=&\prod_{i=1}^{N_f}\exp\{Tr\ln (\gamma_\mu \nabla_\mu +m)\}=\exp W_{E},
\label{2}
\end{eqnarray}

with $W_{E}$ being the Euclidean effective action
\begin{eqnarray}
W_E&=&\sum_{i=1}^{N_f}\int \frac{dp_D}{2\pi }\sum_{k,\kappa}\ln
(p_D^2+\varepsilon_k^2(i,\kappa)),
\label{3}
\end{eqnarray}
and $\varepsilon_k\left( i,\kappa \right) $ is the energy spectrum of
quarks of flavor $i$ and color $\kappa$ with quantum numbers $k$ moving in the
constant external field ${F}_{\mu \nu }^a$.

In the case of finite temperature $T= 1/ \beta >0$ the effective
potential $v=-W_E/(\beta L^{D-1})$ [5] is obtained after substituting $%
p_D\rightarrow \frac{2\pi }\beta (l+\frac 12), l=0,\pm 1,\pm 2,...$,
\begin{eqnarray}
v&=&-\frac 1{\beta L^{D-1}}\sum^{N_f}_{i=1}\sum_{k,\kappa}
\sum^{l=+\infty}_{l=-\infty}\ln\left[\left(\frac{2\pi(l+1/2)}{\beta}\right)^2+
\eps^2_k(i,\kappa)\right]=\nn
&=&\frac 1{\beta L^{D-1}}\sum^{l=+\infty}_{l=-\infty}
\sum^{N_f}_{i=1}\sum_{k,\kappa} \ints^{\infty}_{\frac1{\Lambda^2}}\frac{ds}s
\exp\biggl\{-s\biggl[\left(\frac{2\pi(l+1/2)}{\beta}\right)^2+\nn
&&+\eps^2_k(i,\kappa)\biggr]\biggr\}=
\frac{N_f}{2\sqrt{\pi}L^{D-1}}\sum_{k,\kappa}\ints^{\infty}_{\frac1{\Lambda^2}}
\frac{ds}{s^{3/2}}\e^{\ts-s\eps^2_k(i,\kappa)}\times\label{4}\\
&&\times\left[1+2\sum^{\infty}_{l=1}(-1)^l
\e^{\ts\ts-\frac{\beta^2l^2}{4s}}\right],\nonumber
\end{eqnarray}
where $\Lambda $ is an ultraviolet cutoff ($\Lambda \gg m$). According to
(\ref{2}) we then find for the quark condensate
\begin{eqnarray}
\langle \bar{q}q\rangle &=&
\frac{\ds\int\,d\bar qdq\bar qq\exp\left[-\int\,d^Dx {\cal L} \right]}%
{\ds\int\,d\bar qdq\exp\left[-\int\,d^Dx {\cal L} \right]}=\nn
&=&-\frac1Z\,\frac{\partial Z}{\partial m}=-\frac1{\beta L^{D-1}}
\frac{\ds \partial W_E}{\ds\partial m}=
\frac{\ds \partial v}{\ds\partial m},
\label{5}
\end{eqnarray}
which gives [4]
\begin{eqnarray}
\langle \bar{q}q\rangle &=&
-\frac{mN_f}{L^{D-1}\sqrt{\pi}}\sum_{k,\kappa}\ints^{\infty}_{\frac1{\Lambda^2}}
\frac{\ds ds}{\sqrt{s}}\e^{\ts-s\eps^2_k(i,\kappa)}
\left[1+2\sum^{\infty}_{l=1}(-1)^l \e^{\ts-\frac{\beta^2l^2}{4s}}\right]=\nn
&=&\langle \bar{q}q\rangle_{T=0}+\langle \bar{q}q\rangle_{T\ne0}.
\label{6}
\end{eqnarray}

Clearly, in the case of a vanishing external field ($F_{\mu \nu }^a=0$), we have
$\varepsilon_k^2=\sum_{i=1}^{D-1}k_i^2+m^2$ $(-\infty <k_i<\infty )$.
Then, at $T=0$ one obtains
\begin{eqnarray}
\langle \bar{q}q\rangle &=&-\frac{\ds 3mN_f}{\ds 2^{D-2}\pi^{D/2}}
\ints^{\infty}_{\frac1{\Lambda^2}}\frac{\ds ds}{\ds s^{D/2}}\e^{\ts-s m^2}.
\label{7}
\end{eqnarray}

In the following, we shall analyze three special cases of external chromomagnetic
fields.

{\bf Case $i$):}

Rotational--symmetric non--abelian chromomagnetic field
\begin{eqnarray}
A_1^1&=&A_2^2=A_3^3=\sqrt{\frac Hg},\quad H_i^a=\delta _i^aH (i=1,2,3),
\label{8}
\end{eqnarray}
with all other components of $A_\mu ^a$ vanishing.

The energy spectrum has six branches, two of which correspond to quarks that
do not interact with the chromomagnetic field
\begin{eqnarray}
\varepsilon _{1,2}^2&=&\overrightarrow{p}^2+m^2,
\label{9}
\end{eqnarray}
and the other four are given as follows
\begin{eqnarray}
\varepsilon_{3,4}^2 &=& m^2+(\sqrt{a} \pm \sqrt{{\vec p}^2})^2,\nonumber\\
\varepsilon_{5,6}^2 &=& m^2+(\sqrt{a} \pm \sqrt{4a +{\vec p}^2})^2,
\label{10}
\end{eqnarray}
where $a=gH/4$.

{\bf Case $ii$):}

Axial--symmetric non--abelian chromomagnetic field
\begin{eqnarray}
A_1^1&=&A_2^2=\sqrt{\frac Hg}, H_i^a=\delta _3^a\delta _{i3}H,
\label{11}
\end{eqnarray}
with all other components of the potential vanishing.

The branches of the quark energy spectrum  are besides (\ref{9}) as follows
\begin{eqnarray}
\varepsilon_{3,4,5,6}^2&=&m^2+2a \pm \sqrt{4a^2+4a
p^2_{\perp}}+p^2_3+p^2_{\perp}=\nn
&=&m^2 +p^2_3+(\sqrt{ a +p^2_{\perp}}\pm \sqrt{a})^2.
\label{12}
\end{eqnarray}

{\bf Case $iii$):}

Abelian chromomagnetic field
\begin{eqnarray}
A_\mu ^a&=&\delta^a_3\delta_{\mu2}x_1H.
\label{13}
\end{eqnarray}

This time only two color degrees of freedom of quarks with charges $\pm g/2$
interact with the external field. The energy spectrum of quarks is now given by
\begin{eqnarray}
\varepsilon_{3,4,5,6}^2&=&\eps^2_{n,\sigma,p_3}=gH(n+\frac12+\frac{\sigma}2)+
p^2_3+m^2,
\label{14}
\end{eqnarray}
where $\sigma =\pm 1$ is the spin projection on the external field
direction, $p_3$ is the longitudinal component of the quark momentum ($-\infty <p_3<\infty $),
\begin{eqnarray}
p_{\perp}^2&=&gH(n+\frac12)
\label{15}
\end{eqnarray}
is the transversal component squared of the quark momentum, and $n=0,1,2,...$
is the Landau quantum number.

As can be expected from (\ref{6}), the form of the spectrum is essential for the quark
condensate formation. Using the above three expressions of energy spectra
for field configurations $i$), $ii$) and $iii$), we shall next study the corresponding
three types of quark condensates in the strong field limit.
\subsection*{2.2. Asymptotic estimates for strong fields $\frac{gH}{m^2}\gg
1. $}

{\bf Case $i$):}

According to (\ref{9}), (\ref{10}) we have
\begin{eqnarray}
\langle \bar{q}q\rangle &=& -\frac{\ds m N_f 4\pi a}{\ds \sqrt{\pi}(2\pi)^3}
\ints^{\infty}_{\frac{\ts a}{\ts \Lambda^2}}\frac{\ds dt}{\ds \sqrt{t}}
\e^{\ts-\frac{tm^2}a}\ints^{\infty}_0\,dxx^2
\Bigl[2\e^{\ts-x^2t}+\e^{\ts-t(1-x)^2}+\nn
&&+\e^{\ts-t(1+x)^2}
+\e^{\ts-(1+\sqrt{x^2+4})^2t}+\label{16}\\
&&+\e^{\ts-(1-\sqrt{x^2+4})^2t}\Bigr]\left(1+2\sum^{\infty}_{l=1}(-1)^l \e^{\ts-
\frac{\beta^2l^2a}{4t}}
\right).\nonumber
\end{eqnarray}

Taking the $T=0$ term in (\ref{16}),
we see that the branch of the spectrum
\[
\varepsilon _4^2= m^2+(\sqrt{a}-\sqrt{{\vec p}^2})^2
\]
plays the main role when $m\rightarrow 0$ (the second term in the brackets).
When $h=gH/m^2\gg 1$ the following asymptotics is obtained
\begin{eqnarray}
\langle \bar{q}q\rangle &=& -\frac{m^3N_f}{4\pi^2}
\left[3\left(\frac{\Lambda^2}{m^2}-\ln\frac{\Lambda^2}{m^2}\right)
+\frac h2\ln(C_{1}h)-hI_1(\beta m)\right].
\label{17}
\end{eqnarray}

Here
\begin{eqnarray*}
I_1(\beta m)&=& -\sum^{\infty}_{l=1}(-1)^l \ints^{\infty}_0\frac{dx}x
\exp\left[-\left(x+\frac{l^2 m^2 \beta^2}{4x}\right)\right]=\nn
&=&-2\sum^{\infty}_{l=1}(-1)^l K_0(\beta ml ),
\end{eqnarray*}
where $K_0(y)$ is the Macdonald`s function and $C_1$ is a certain numerical constant.

It is well-known that $D\chi SB$, signalled by a
nonvanishing quark condensate, is the origin of dynamical quark masses. The
underlying mechanism can be most simply illustrated by considering an NJL model
with four-fermion interactions,
\begin{eqnarray}
{\cal L}_{int}&=&2G\sum^{N^2_f-1}_{i=0}\left\{(\bar q\frac12\lambda^F_iq)^2+
(\bar qi\gamma^5 \frac12\lambda^F_iq)^2\right\},\nn
&&\Tr \lambda^F_i \lambda^F_j=2\delta_{ij}, \quad \lambda^F_0=\sqrt{2/N_f}{\bf 1},
\label{18}
\end{eqnarray}
with $G$ being a universal coupling constant and $\lambda^F_i$ being flavor generators.

After applying the bosonization procedure in the $N_c\rightarrow \infty $ limit
($N_cG=const$), the one--loop expression for the corresponding meson functional integral is dominated
by the
stationary phase, and we obtain the gap equation~[7]\footnote{The limit $N_c\rightarrow \infty $
is here only needed for technical reasons. In subsequent applications, one can set
afterwards again $N_c=3$ in final expressions.}

\begin{eqnarray}
m&=&-\frac{2G}{N_f}\langle \bar{q}q\rangle,
\label{19}
\end{eqnarray}
or, according to (\ref{17}):
\begin{eqnarray}
\Lambda ^2(\frac 1{\widetilde{g}}-1)&=& -m^2 \ln\frac{\Lambda^2}{m^2}+
m^2 \frac h6 \ln C_1h- h\frac{m^2}3I_1,
\label{20}
\end{eqnarray}
where $\widetilde{g}=\frac{\ts 3\Lambda ^2G}{\ts 2\pi^2}$.

For $gH\ln\frac{gH}{m^2}\gg m^2 \ln\frac{\Lambda^2}{m^2}$ $(gH< \Lambda^2)$
we have  a  solution of (\ref{20}) even for weak coupling $\tilde g\ll 1$
\begin{eqnarray}
m^2(T)&=&C_1gH\exp\left[-\frac{4\pi^2}{GgH}-2I_1(\beta m(T))\right].
\label{21}
\end{eqnarray}

In particular, for $T=0$,
\begin{eqnarray}
m^2(0)&=&C_1gH\exp\left(-\frac{4\pi^2}{GgH}\right).
\label{22}
\end{eqnarray}

The critical temperature can now be found from the condition $m(T_C)=0$,
which gives
\begin{eqnarray}
T_C&=&\pi^{-1}\e^{\ts\gamma} m(0)\simeq 0,5669\, m(0),
\label{23}
\end{eqnarray}
where $\gamma$ is Euler's constant.
Let us emphasize that the result (\ref{17}) with the logarithmic term $\frac h2\ln h$ demonstrates
the effect of dimensional reduction $D=3+1\rightarrow D=1+1$. Indeed, integration of the main term
in (\ref{16}) gives
\begin{eqnarray}
\langle \bar{q}q\rangle & \simeq & -\frac{m N_f a}{2\pi^2}
\ints^{\infty}_{\frac1a} \frac{ds}s \e^{\ts-s m^2} \approx
-\frac{m N_f a}{2\pi^2} \ln\frac{a}{m^2},
\label{24}
\end{eqnarray}
which corresponds to (\ref{7}) with $D=2$ and $\Lambda^2$ replaced by $a$.

{\bf Case $ii$:}

In this case we have
\begin{eqnarray*}
\langle \bar{q}q\rangle &=&-\frac {m N_f}{2\pi^2}
\ints^{\infty}_{\frac1{\Lambda^2}} \frac{ds}s \e^{\ts-s m^2}
\ints^{\infty}_0\,dp_{\perp}p_{\perp}\times\\
&&\times\left[\e^{\ts-sp^2_{\perp}}+
\e^{\ts-s(\sqrt{a+p^2_{\perp}}-\sqrt{a})^2}+\right.\\
&&\left.+\e^{\ts-s(\sqrt{a+p^2_{\perp}}+\sqrt{a})^2}\right]
\left[1+2\sum^{\infty}_{l=1}(-1)^l \e^{\ts-\frac{\beta^2l^2}{4s}}\right].
\end{eqnarray*}

The gap equation for $h\gg 1$ now takes the form
\begin{eqnarray}
\Lambda^2\left(\frac1{\tilde g}-1\right)&=&-m^2 \left(\ln\frac{\Lambda^2}{m^2}-
\frac h3 -\frac{\sqrt{\pi h}}3I_2(\beta m)\right),
\label{25}
\end{eqnarray}
where
\begin{eqnarray*}
I_2(\beta m)&=&\sum^{\infty}_{l=1}(-1)^l \ints^{\infty}_0\frac{dx}{x^{3/2}}
\exp\left[-\left(x+\frac{\beta^2l^2m^2}{4x}\right)\right]=\\
&=&2\sqrt{\frac{\pi}{\beta m}}\sum^{\infty}_{l=1}(-1)^l
\frac{\e^{\ts-\sqrt{\beta m l}}}{\sqrt{l}}.
\end{eqnarray*}

Note that the $T \rightarrow\ 0$ limit of expression (\ref{25}) differs from the corresponding one in [13] by the sign of the term $h/3$. A nontrivial solution of (\ref{25}) is then possible for
\[
\tilde g> \frac1{1+(gH/(3\Lambda^2))}.
\]
and it exists even for $\tilde g<1$, i.e. for weak quark attraction. It has the form
\begin{eqnarray}
m^2&=&\Lambda^2 \exp\left[-\frac{\Lambda^2}{m^2}\left(1-\frac1{\tilde g}\right)-
\frac h3-\frac{\sqrt{\pi h}}3I_2(\beta m)\right],
\label{26}
\end{eqnarray}
and demonstrates the possibility of chiral symmetry breaking in a non--abelian chromomagnetic
field at $D=3+1$ for $\tilde g<1$ (weak attraction).

The dependence on $h$ in (\ref{25}) is found from the dominating
term in (\ref{6}) arising from the branch
\[
\eps^2=m^2+p^2_3+(\sqrt{p^2_{\perp}+a }-\sqrt{a })^2 .
\]
Then we have for $a \rightarrow\infty$
\[
\langle \bar{q}q\rangle \sim \ints^{\infty}_{\frac1a}\frac{ds}s\e^{\ts-s m^2}
\ints^{\infty}_0\,dp_{\perp}p_{\perp}
\e^{\ts-\frac{\ts sp^4_{\perp}}{\ts 4a}}  \sim
\sqrt{a} \ints^{\infty}_{\frac1a} \frac{ds}{s^{3/2}} \sim a
\]
corresponding to (\ref{7}) with $D=3$, which demonstrates the $3+1 \rightarrow 2+1$
dimensional reduction in this type of field.

{\bf Case $iii$:}

For the abelian chromomagnetic field with the spectrum (\ref{14}) we obtain
\begin{eqnarray}
\langle \bar{q}q\rangle &=&-\frac {m^3N_f}{4 \pi^2}
\Bigl\{h\ln\frac h{2\pi}+3\left(\frac{\Lambda^2}{m^2} -
\ln\frac{\Lambda^2}{m^2}\right)- 2hI_1(\beta m)\Bigr\},
\label{26}
\end{eqnarray}
which is similar to (\ref{17}), but differs by an overall factor 2 in field--dependent
terms. This difference is simply due to the fact, that the main term $h\ln h$ is
obtained from two colors in the spectrum (\ref{14}), while in the non--abelian case
only one branch of the spectrum contributes to (\ref{17}).

For $gH\ln\left(\frac{gH}{m^2}\right) \gg
m^2 \ln\left(\frac{\Lambda^2}{m^2}\right)$\,$(gH< \Lambda^2)$
we obtain for $m^2(T)$, $m^2(0)$ and $T_C$ the same equations
(\ref{21})--(\ref{23}) as in the non--abelian case $i)$, but with the obvious
replacements $C_1 \rightarrow 1/2\pi$ and $4 \pi^2\rightarrow 2\pi^2$ in the
exponents.

The main logarithmic term in (\ref{26}) is obtained from the $n=0$, $\sigma=-1$
contribution in the sum over quantum states in (\ref{6})
\begin{eqnarray*}
\langle \bar{q}q\rangle & \sim & -\ints^{\infty}_{\frac1{\Lambda^2}}
\frac{ds}{\sqrt{s}}\ints^{+\infty}_{-\infty}\,dp_3 \sum^{\infty}_{n=0}
(2-\delta_{n0})\exp[-gHns-s m^2 -p^2_3 s]\sim \\
& \sim & -\ints^{\infty}_{1/gH}\frac{ds}s \e^{\ts-s m^2} \approx
-\ln\frac{gH}{m^2}.
\end{eqnarray*}
Obviously, this corresponds to (\ref{7}) with $D=2$, which demonstrates the dimensional reduction
in this case $3+1\rightarrow 1+1$, similar to the non--abelian case $i)$.
(The replacement $\Lambda^2 \rightarrow gH$ follows here from the requirement
$1/\Lambda^2 < 1/gH \ll s$ for the integration region.)

\section*{3. Summary and discussions}

As shown in this letter, the phenomenon of $D\chi SB$  does exist
for all the field configurations considered here even for weak
coupling. This effect
is accompanied by an effective lowering of dimensionality in strong
chromomagnetic fields, where the number of reduced units of dimensions depends
on the concrete type of the field. It should also be noted that there is no contradiction
with the Mermin--Wagner--Coleman (MWC) theorem [15], when the effective
dimensionality reduces to the value $1+1$, since only charged channels are
affected by this reduction, while Nambu--Goldstone bosons are neutral [12].
It is interesting to mention that our result (\ref{21}), being effectively $1+1$
dimensional, corresponds to considerations of paper [16], where a similar
expression for the dynamical mass has been obtained directly in the $1+1$
dimensional Gross--Neveu model.

The physical reason of dimensional reduction is quite transparent on the basis of
the symmetry properties of the external fields and the corresponding degeneracy
of the quark spectra. If the field potentials are rotationally symmetric as
in (\ref{8}), the spectrum depends on $|\vec p|$, being twice degenerated
with respect to azimuthal and polar angles of the momentum vector
$\vec p$. The branch of the spectrum
$\eps^2_4=m^2+(\sqrt{a}-\sqrt{\vec p^2})^2$ providing the main
contribution to the condensate formation (\ref{17}) has its minimum
$\eps^2_{min}=m^2$ at nonzero value of $|\vec p|=\sqrt{a}$. In this
case the dimensionality is lowered from
$3+1$ to $1+1$, i.e. by 2 units, which is equal to the degeneracy of
the quark energy spectrum. For the axially symmetric potential (\ref{11})
the spectrum of quarks is a function of longitudinal $p_3$
and transversal $p_{\perp}$
components of the momentum vector $\vec p$, while $\eps^2=\eps^2_{min}=m^2$
for $p_{\perp}=p_3=0$. Now dimensions are
reduced by 1 unit $(3+1\rightarrow 2+1)$ which is again equal to the
degeneracy of the quark spectrum. Finally, for the abelian potential (\ref{13})
the main contribution for strong fields comes from the small
region around the minimum value $\eps^2_{min}=m^2$  for $p_3=0$,
$n=0$, $\sigma=-1$. In this case $(p^2_{\perp})_{min}=gH/2\gg m^2$
and a quark is localized in a small region $(\bigtriangleup r)^2_{min} \sim
2/(gH) \ll \frac{\ts 1}{m^2}$ in the $xy$--plane which clearly demonstrates
$3+1\rightarrow 1+1$ effective dimensional reduction by 2 units.

We note in passing that the Dirac equation in both the abelian and non--abelian
chromomagnetic fields possesses the property of supersymmetry [14], which
provides for the lowest possible value of the energy $\eps^2_{min}=m^2$.
This is crucial for the formation of the quark condensate in the infrared
region $m \rightarrow 0$ $(gH\gg m^2)$ in the strong field limit.
In the opposite case of weak chromomagnetic fields, $gH \ll m^2$, the mass
generation is possible only in the strong coupling regime, when $\tilde g>1$.
In this case the gluon condensate has a stabilizing effect in contrast to the strong field case.
Here the dynamical mass $m$ and $T_C$ increase with the field [8,9].

Finally, we remark that it is interesting to study also the effect of the
temperature dependence of the external chromomagnetic field [5] not only
in the weak  field limit [9] but also for the case of strong fields.

\section*{Acknowledgements}
One of the authors (V.Ch.Zh.) gratefully acknowledges the hospitality of
Prof. Mueller-Preussker and his colleagues at the particle theory group of
the Humboldt University extended to him during his stay there.

\end{document}